\documentclass{svmult}

\usepackage{makeidx}         
\usepackage{graphicx}        
\usepackage{multicol}        
\usepackage[bottom]{footmisc}

\makeindex             

\begin{document}

\title*{The Power (Law) of Indian Markets:\\
{\Large Analysing NSE and BSE trading statistics}}
\titlerunning{The Power (Law) of Indian Markets} 
\author{Sitabhra Sinha and Raj Kumar Pan}
\institute{The Institute of Mathematical Sciences, C. I. T. Campus,
Taramani,\\ 
Chennai - 600 113, India.\\
\texttt{sitabhra@imsc.res.in}}

\maketitle
The nature of fluctuations in the Indian financial market is analyzed in 
this paper. 
We have looked at
the price returns of individual stocks, with tick-by-tick data from the 
National Stock Exchange (NSE) and daily closing price data from both NSE
and the Bombay Stock Exchange (BSE), the two largest exchanges in India. We find
that the price returns in Indian markets follow a fat-tailed cumulative 
distribution, consistent with a power law having exponent $\alpha \sim 3$,
similar to that observed in developed markets. However,
the distributions of trading volume and the number of trades have a different
nature than that seen in the New York Stock Exchange (NYSE). Further,
the price movement of different stocks are highly correlated in Indian
markets.
\section{Introduction}
\label{sec:1}
\vspace{-0.25cm}
Over the past decade, a growing number of physicists have got involved
in searching for statistical regularities in the behavior of financial
markets. A key motivation for such ``econophysicists" is the prospect
of discovering {\em universal} features in financial data \cite{farmer05},
i.e., statistical properties that are invariant with respect to stocks,
markets, the time interval over which the data is collected, etc.
The most prominent candidate for such universality is the distribution
of fluctuations in the price of individual stocks \cite{lux96,plerou99}, 
as well 
as, market indices \cite{gopikrishnan99} which reflect the composite value 
of many such stocks.
Studies in various markets have reported evidence for the cumulative 
distribution of price fluctuations having positive and negative tails 
that obey a power law decay,
i.e., $P_c (x) \sim x^{-\alpha}$.
It has also been claimed that the exponent for this power law, $\alpha$,
is around 3 for most markets (the ``inverse cubic law") \cite{gopikrishnan98}.
It may be useful here to distinguish between the power law reported for
individual stock price fluctuations and that for market index fluctuations,
as the former is more fundamental and implies the latter, provided most
of the stocks comprising the index have significant cross-correlation in
their price movement. We will, therefore, focus on the behavior of
individual stocks, although we will also mention in brief our study of
a particular Indian market index.

The prime motivation for our study of the Indian financial market is
to check recent claims that emerging markets (including those in India) have
behavior that departs significantly from the previously mentioned
``universal" behavior for developed markets.
Although a recent paper \cite{sarma05} reported heavy tailed behavior of 
the fluctuation distribution for an Indian market index between 
Nov 1994 and Oct 2004, 
the generalized Pareto distribution fit to the data
did not suggest a power law decay of the tails.
Moreover, an almost contemporaneous study \cite{matia04}
of the fluctuations in the
price of 49 largest stocks in the NSE 
between Nov 1994 and Jun 2002,
has claimed that the
distribution has exponentially decaying tails.
This implies the presence of a characteristic scale, and therefore,
the breakdown of universality of the power law tail for the price 
fluctuation distribution. The contradiction between the results of
the two groups indicates that a careful analysis of the Indian market
is necessary to come to a conclusive decision. Note that, both
of the above-mentioned studies looked at low-resolution data,
namely, the daily closing time series. 

In this study, we have
looked at the high-frequency transaction by transaction stock price
data, as well as taken a fresh look at the low-frequency
daily data. We conclude that,
far from being different, the distribution of
price fluctuations in Indian markets is {\em quantitatively} 
almost identical to that of developed markets.
However, the distributions for trading volume and
number of trades seem to be market-specific, with the Indian data being
consistent with a log-normal distribution for both of these quantities. 
Next, we look at the distribution of fluctuations in the NSE market index
and find it to also follow the ``inverse cubic law".
Given the result for the price fluctuation distribution of 
individual stocks, this is expected if the price
movements of the various stocks are highly correlated. Therefore, we
also study the cross-correlations among the price fluctuations 
of most
of the stocks comprising the index.
We find that, on the whole, stock price movements in the Indian market are
remarkably correlated.

\vspace{-0.5cm}
\section{The Indian Financial Market}
\label{sec:2}
\vspace{-0.25cm}
There are 23 different stock markets in India. The two largest are 
the National Stock Exchange (NSE) and the Bombay Stock Exchange (BSE) which
together accounted for more than $98\%$ of the total turnover for
all markets in 2003-04 \cite{ismr}. Of these, the NSE is the larger one, with
a turnover that is slightly more than double that of BSE, although 
their market capitalizations are comparable.
BSE was founded in 1875, and is the oldest
stock market in Asia. It has the largest number of companies listed
and traded, among all the exchanges in India. The market
indices associated with it, namely BSE 30, BSE 100 and BSE 500, are closely
followed indicators of the health of the Indian financial market. The stocks
belonging to BSE 500 represent nearly 93\% of the total market capitalisation 
in that
exchange, and therefore in this study we have confined ourselves to these
stocks.

Compared to BSE, NSE is considerably younger, having
commenced operations in the capital (equities) market from Nov 1994. 
However, as of 2004, it is already the world's third largest stock 
exchange (after NASDAQ and NYSE) in terms of transactions \cite{ismr}. 
The most important market index associated with
the NSE is the Nifty. The 50 stocks comprising the Nifty index
represent about 58\% of the total market capitalization 
and 47\% of the traded value of all stocks in the NSE
(as of Dec 2005).

{\em Description of the data set.} The low-frequency data that we analyze 
consists of the daily
closing price, volume and number of trades for individual stocks from 
BSE (starting from as early as 1991) and NSE (starting from as early as 1994).
This data is available from the web-sites of the corresponding 
exchanges \cite{websites}.  
The high-frequency tick-by-tick data contains information of all
transactions carried out in the NSE between Jan 1, 2003 and Mar 31, 2004.
This information includes the date and time of trade, the price of the
stock during transaction
and the volume of shares traded. This database is available in the
form of CDs published by NSE. For calculating the price return, we have 
focused on 479 stocks, which were all used to calculate the BSE 500 index 
during this period. To calculate the distribution of index fluctuations,
we have looked at the daily closing value of Nifty between Jan 1, 1995
and Dec 31, 2005.
For cross-correlation analysis,
we have focused on daily closing price data of 45 NSE stocks 
(all belonging to the Nifty index) from Jan 1, 1997 to Dec 31, 2005.

\vspace{-0.5cm}
\section{Price Return Distribution of Individual Stocks}
\label{sec:4}
\vspace{-0.25cm}
To measure the price fluctuations (or the fluctuations in
the market index) such that the result is independent of the scale of 
measurement, we calculate the logarithmic return of price (or index). 
If $P_{i}(t)$ is the stock price of the $i$th stock at time $t$, then the
(logarithmic) price return is defined as
\begin{equation}
R_{i}(t,\Delta t) \equiv \ln {P_{i}(t+\Delta t)}- \ln {P_{i}(t)}. \label{return}
\end{equation}
However, the distribution of price returns of different stocks may
have different widths, owing to differences in their volatility. 
To be able to compare the distribution of various
stocks, we must normalize the returns by dividing them with their
standard deviation (which is a measure of the volatility),
$ \sigma_{i} = \sqrt{\langle R_{i}^{2} \rangle - \langle R_{i} \rangle^{2}}.$
The normalized price return is, therefore, given by
\begin{equation}
r_{i}(t,\Delta t) \equiv \frac{R_{i}-\langle R_{i} \rangle}{\sigma_{i}},
\label{normalized_return}
\end{equation}
where $\langle \ldots \rangle$ represents time average.

\subsection{Daily Price Returns in BSE and NSE}
\begin{figure}[tbp]
\centering
\includegraphics[width=0.49\linewidth,clip]{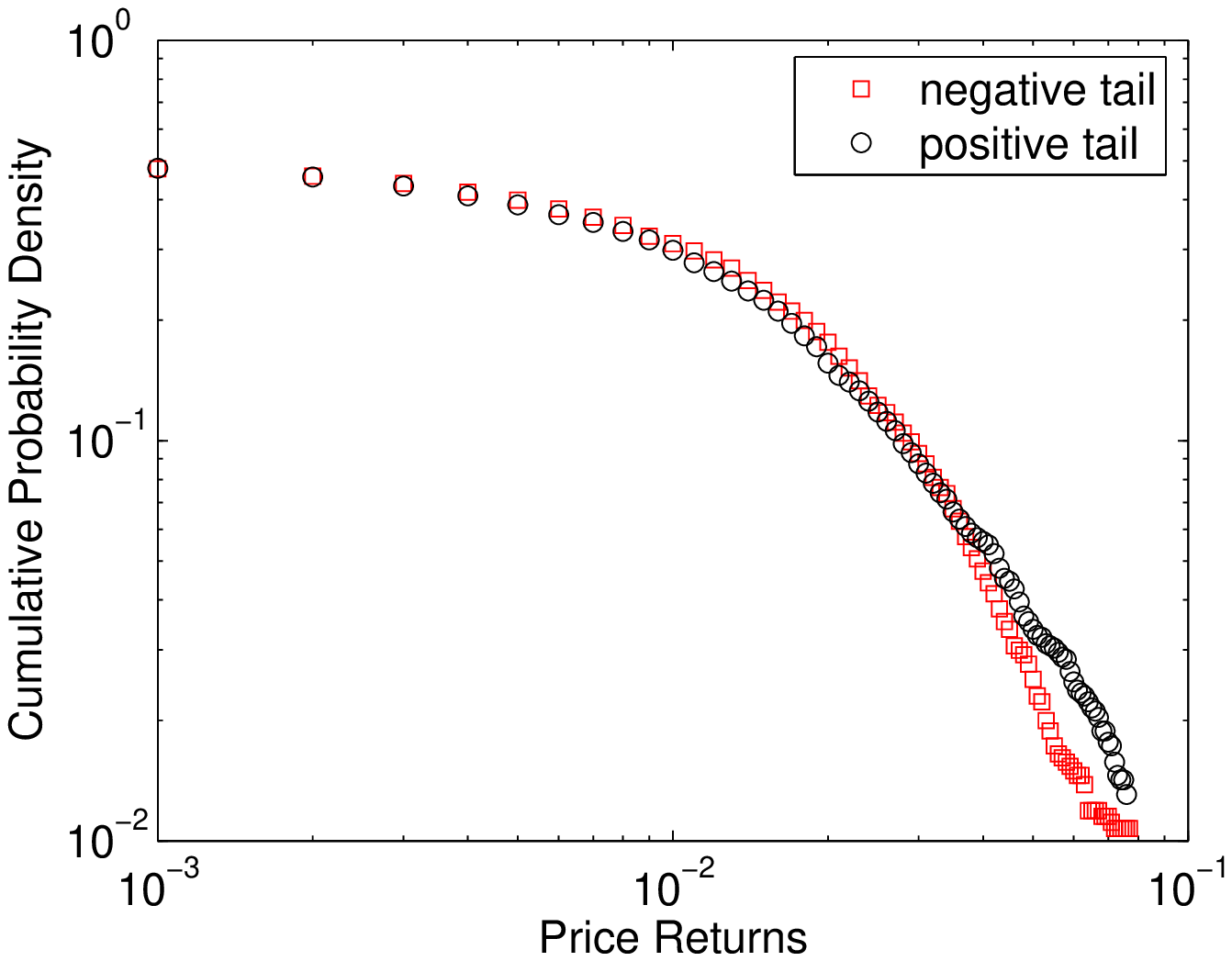}
\includegraphics[width=0.49\linewidth,clip]{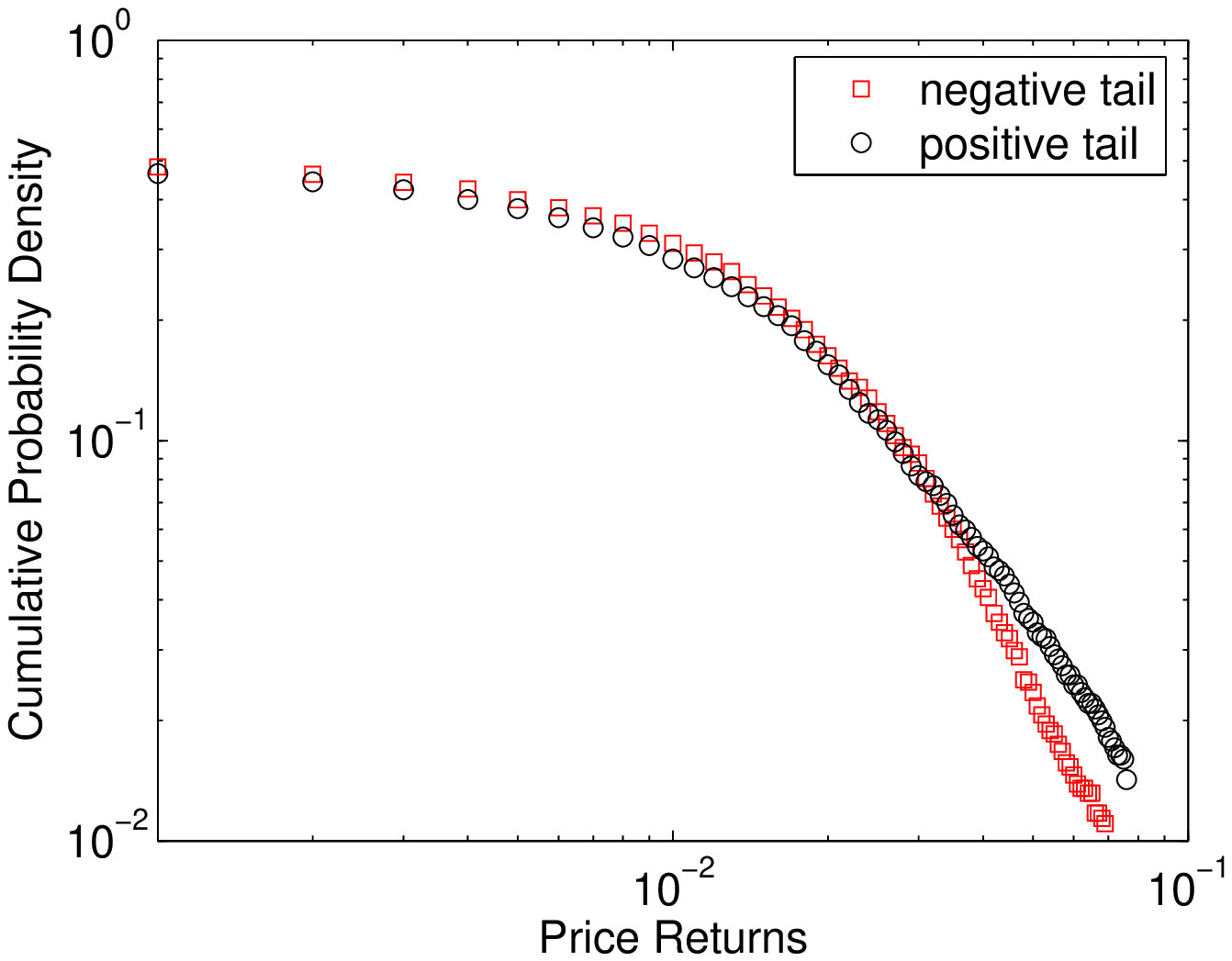}\\
\caption{Cumulative distribution of daily price returns for a 
particular stock (Reliance) at BSE (left) between July 12, 1995 and
Jan 31, 2006, and at NSE (right) between Nov 3, 1994 and Jan 30, 2006.}
\label{ss:nsebsedaily}       
\vspace{-0.15cm}
\end{figure}
We start by focussing on the daily price variation of individual stocks,
i.e., $\Delta t$ = 1 day.
By using the time series of daily closing price of a particular stock 
(e.g., Reliance) we can obtain the corresponding daily returns. Binning
this data appropriately we can obtain the probability density 
function, and by integrating it over a suitable range, the cumulative 
distribution function (CDF), which is 
essentially the probability that a return is larger than a given value.
Fig.~\ref{ss:nsebsedaily} shows the CDF for daily price returns for the
same stock in BSE (left) and NSE (right). Note that,
we have shown the tails for the 
positive and negative returns in the same figure. The distribution for the
two exchanges are almost identical, and both show long tails consistent
with a power law decay. 

\begin{figure}[tbp]
\centering
\includegraphics[width=0.49\linewidth,clip]{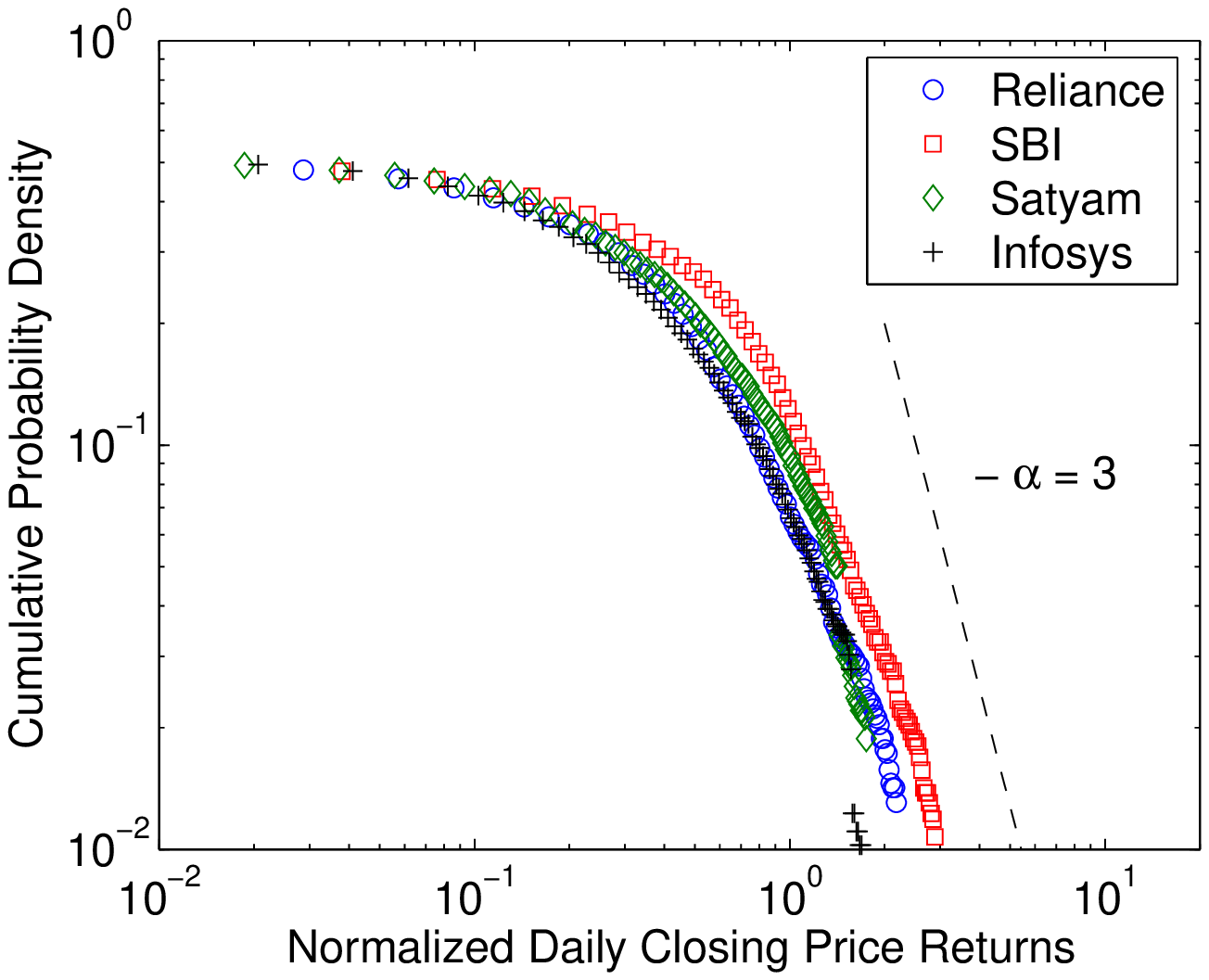}
\includegraphics[width=0.49\linewidth,clip]{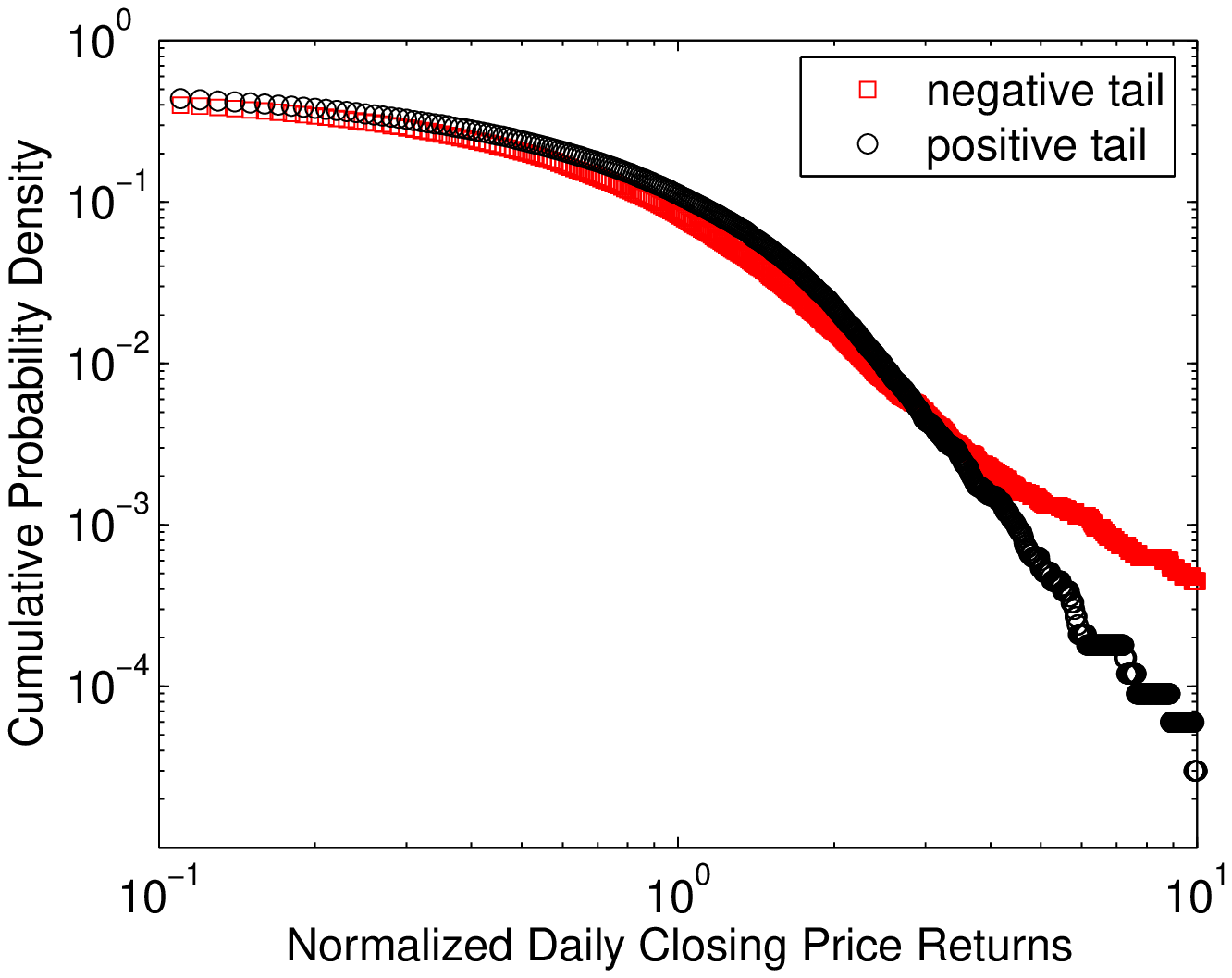}\\
\caption{(Left) Cumulative distribution of the normalized daily closing price 
returns for four stocks in BSE between July 12, 1995 and
Jan 31, 2006. (Right) Cumulative distribution of the aggregated normalized daily
closing price returns for 43 stocks (included in the Nifty index) at NSE 
between Jan 1, 2003 and Jan 30, 2006.}
\label{ss:compare}       
\vspace{-0.15cm}
\end{figure}

To confirm that this is a general property, and not unique to the particular
stock that is being analysed, we next perform the same analysis for other
stocks. To be able to compare between stocks, we normalize the returns for 
each stock by their standard deviation.  
Fig.~\ref{ss:compare}~(left) shows that four stocks chosen
from different sectors have very similar normalized cumulative distributions. 
Moreover, the tail of each of these distributions approximately follow a power
law with exponent $\alpha \simeq 3$. However, the daily closing price data set 
for any particular stock that we have analyzed is not large enough for an 
unambiguous determination of the nature of the tail. For this, we aggregate
the data for 43 frequently traded stocks, all of which are used for calculating 
the Nifty index, over 3 years, and obtain the corresponding CDF 
(Fig.~\ref{ss:compare}, right). Putting together the time series of different
stocks to form a single large time series is justified because, after 
normalization, the different stocks have almost identical distributions 
\cite{plerou99}. From this figure we confirm that the
distribution does indeed follow a power law decay, albeit
with different exponents for the positive and negative return tails.
The different exponents of the positive and negative tails have also
been observed in the case of stocks listed in the New York Stock Exchange
(NYSE) \cite{plerou99}. 
For comparison, we carried out a similar study
with the daily closing price series of several NYSE stocks from
Yahoo! Finance \cite{Yahoo}, and obtained distributions qualitatively similar
to that shown here for the Indian market.

\subsection{Price Return for Tick-by-Tick Data in NSE}
 \begin{figure}[tbp] \centering
\includegraphics[width=0.49\linewidth,clip]{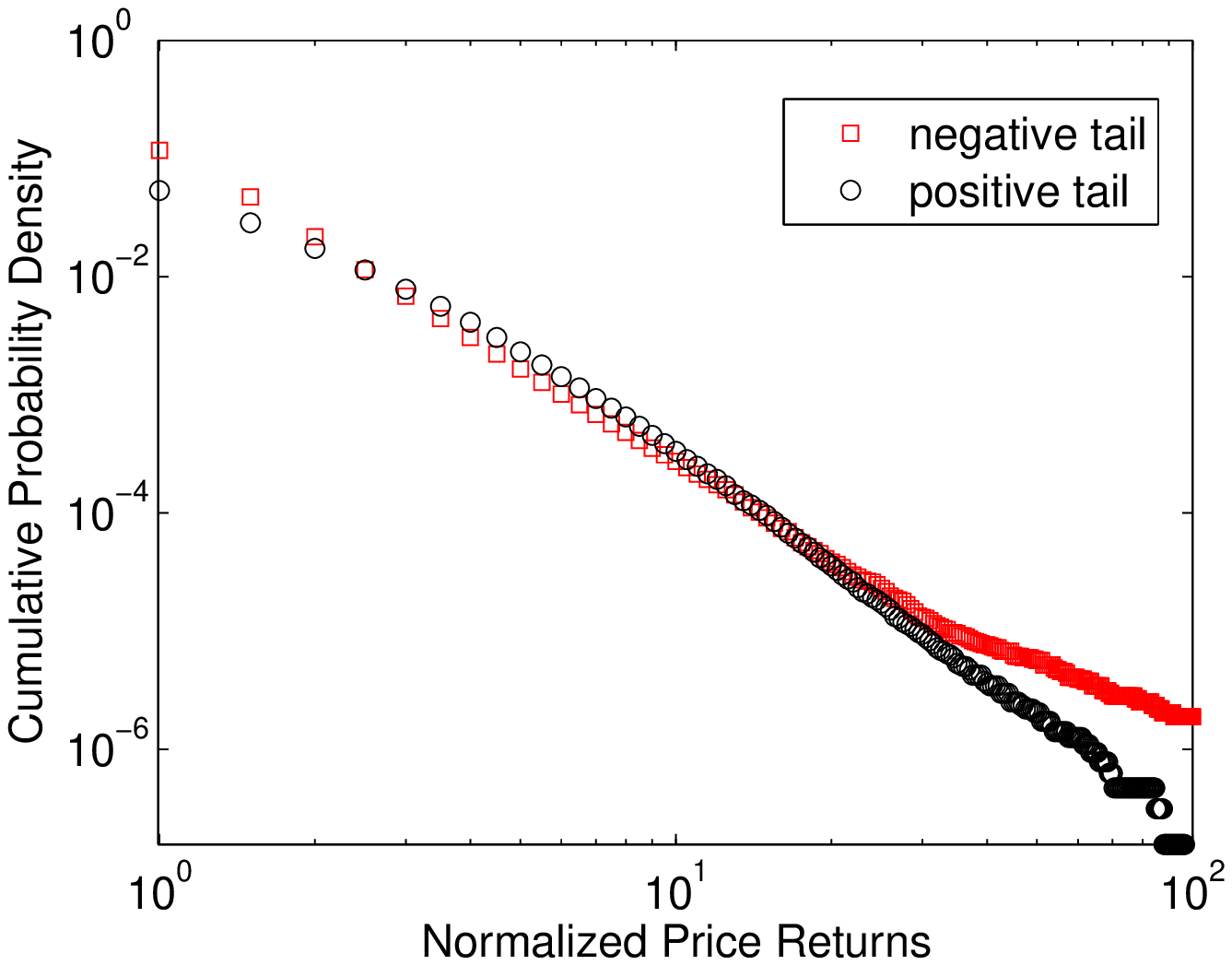}
\includegraphics[width=0.49\linewidth,clip]{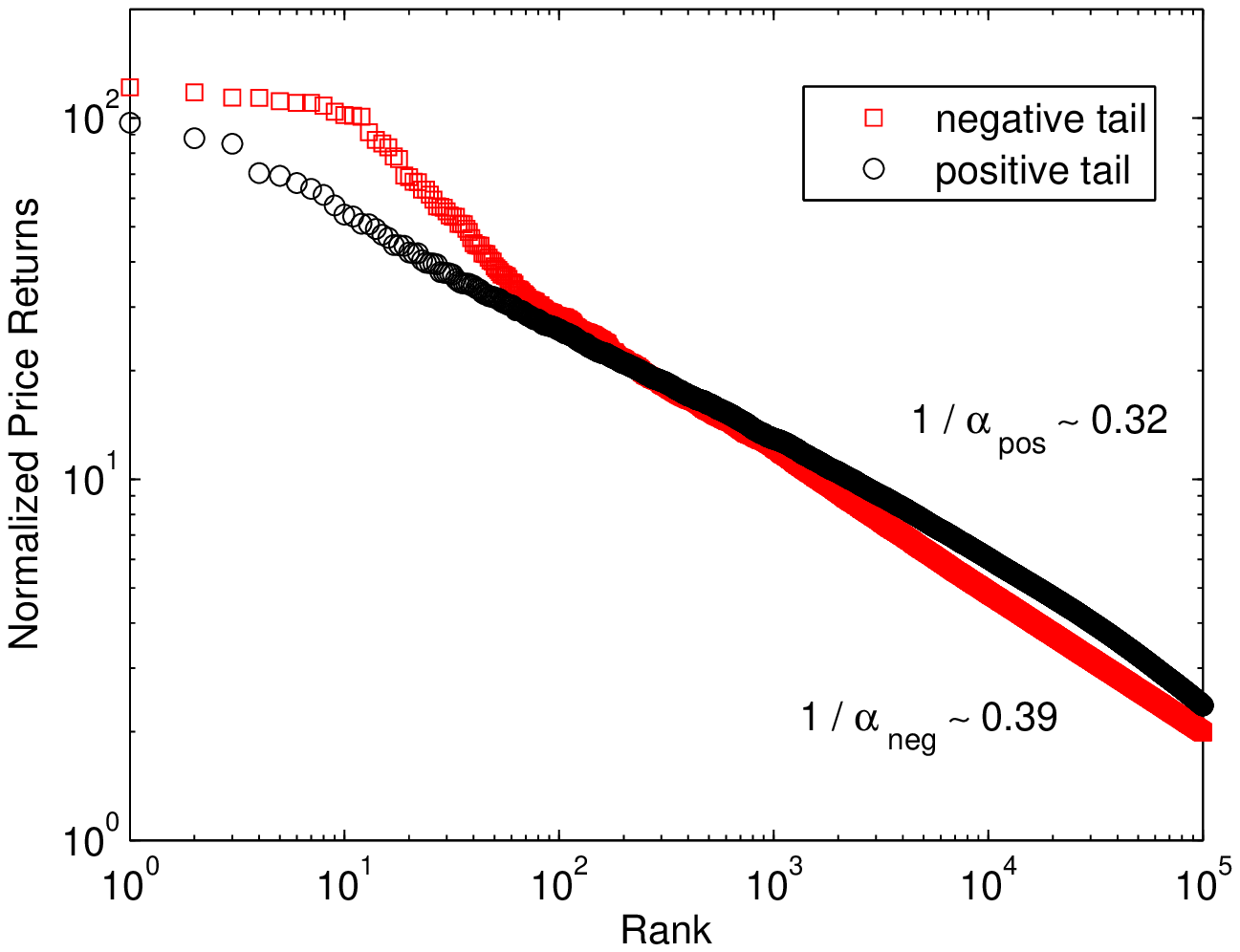}
\caption{Cumulative distribution (left) and rank-ordered plot (right) for
the 5-minute interval price returns aggregated for 479 stocks at NSE
between Jan 1, 2003 to Mar 31, 2004.}
\label{ss:nse479}       
\vspace{-0.15cm}
\end{figure}
The daily data is strongly suggestive of a power
law tail for the price return distribution, but for conclusive evidence
we next turn to the tick-by-tick data for stocks listed in the NSE.
Choosing an appropriate $\Delta t$, we obtain the corresponding return
by taking the log ratio of consecutive average prices, averaged over
a time window of length $\Delta t$;
for the results reported here $\Delta t = 5$ minutes.
We have verified that the nature of the distribution is not sensitive
to the exact value of this parameter.
For individual stocks, the cumulative distribution of returns again show
power law decay, but as the data set for each stock is not large enough,
we carry out an aggregation procedure similar to that outlined above.
Picking 479 frequently traded stocks from NSE,
we put together their normalized returns to form a 
single large data set. The corresponding CDF is shown in Fig.~\ref{ss:nse479}
(left), with the exponents for the positive and negative tails estimated
to be $\alpha \sim 3.2$ and 2.7, respectively. To check the accuracy of these
exponents, obtained using linear least square fitting on a doubly logarithmic
plot, we next plot the return data in descending order. 
This {\em rank-ordered plot}
is an alternative visualization of the CDF, interchanging the ordinate
and abscissae. It is easy to show that if the CDF has a power-law form, so
does the rank-ordered plot, and the two exponents are the inverses of each
other \cite{adamic}. 
Exponents obtained by least square fitting on this graph produces similar values
of $\alpha$, namely, 3.1 and 2.6 for the positive and 
negative tails, respectively.

\subsection{The ``Inverse Cubic Law" for Price and Index Fluctuations}
\begin{figure}[tbp] \centering
\includegraphics[width=0.49\linewidth,clip]{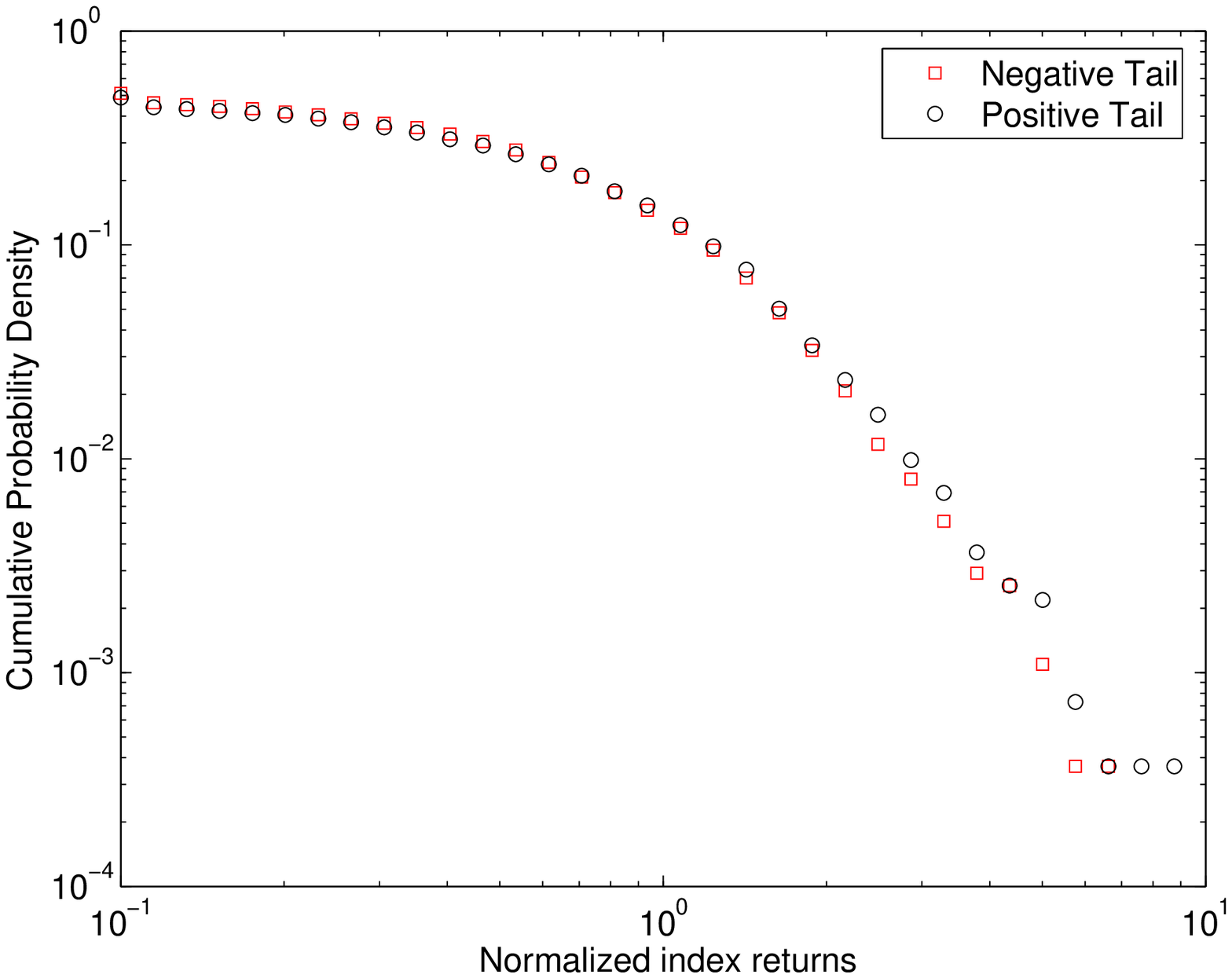}
\includegraphics[width=0.49\linewidth,clip]{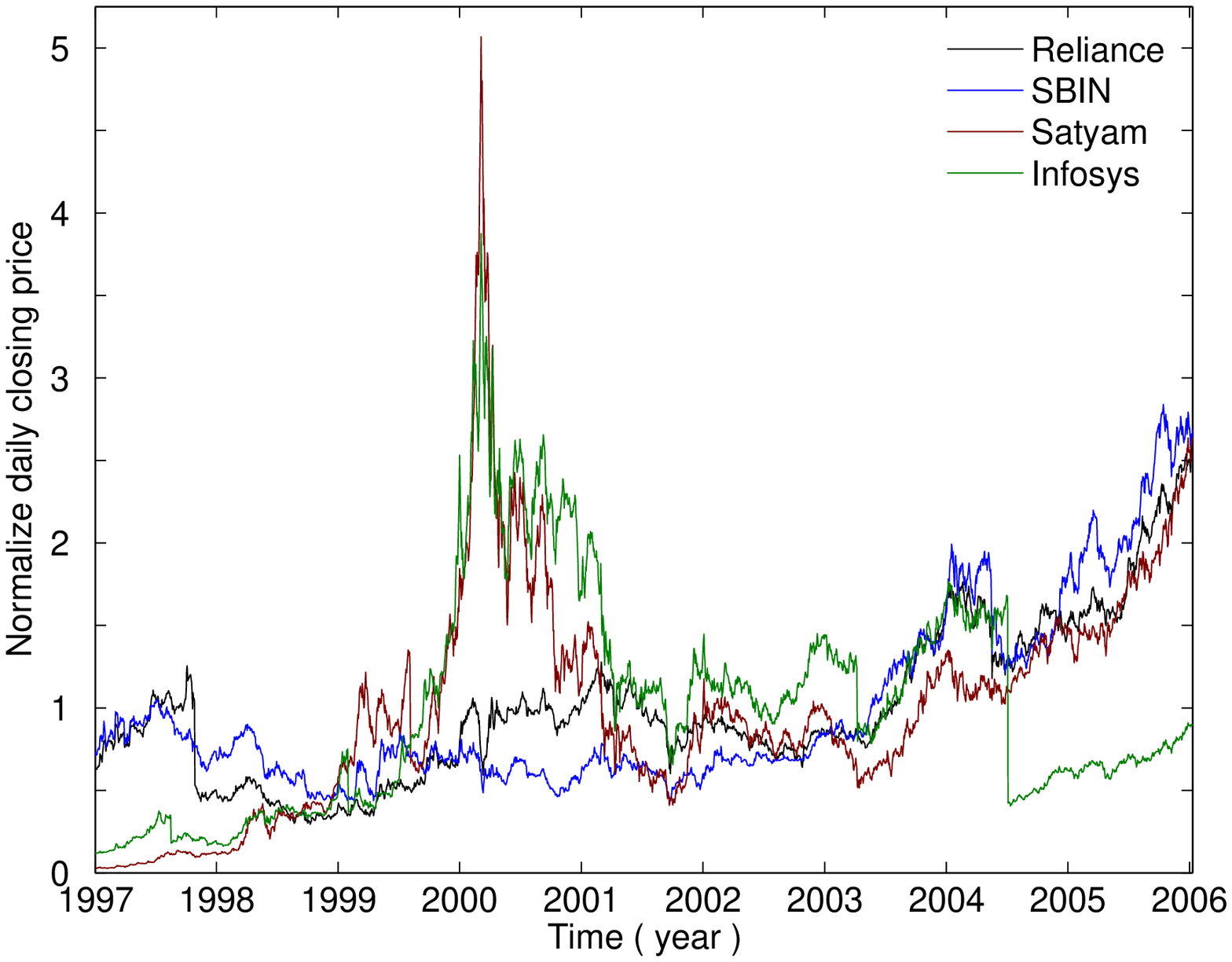}
\caption{
(Left) Cumulative distribution of daily returns
for the Nifty index between Jan 1, 1995 and Dec 31, 2005.
(Right) Comparison of the daily closing price for four stocks in NSE
from Jan 1, 1997 to Dec 31, 2005, showing the high degree of correlation
among the stocks. 
}
\label{nifty}       
\vspace{-0.15cm}
\end{figure}
The results reported above provide conclusive evidence that the
Indian financial market follows a price fluctuation distribution
with long tails described by a power law. Moreover, the exponent
characterizing this power law is close to 3, as has been observed
for several financial markets of developed economies, most
notably the NYSE, where the ``inverse cubic law'' has been found to be 
valid from $\Delta t=$ 1 day to 1 month. 

Most observations of this ``law'' have been in the context of
market indices, rather than the price of individual
stocks. We have, therefore, carried out a similar analysis for the
Nifty index of NSE during the period Jan 1, 1995 to Dec 31, 2005. 
Fig.~\ref{nifty}~(left) shows that the distribution of index returns
also shows a power law decay, with an exponent very close to 3.
As the index is a composite of several stocks, this behavior
can be understood as a consequence of the power law decay
for the tails of individual stock price returns, provided the movement
of these stocks are correlated. As is evident from Fig~\ref{nifty}~(right),
this condition is indeed satisfied in the Indian market.
In a later section we provide a more detailed look into the
cross-correlation structure of these price fluctuations.

These findings assume importance in view of the recent claims that
emerging markets behave very differently from developed 
markets, in particular, exhibiting an exponentially decaying return
distribution \cite{matia04}. India is one of the largest
emerging markets, and our analysis of the price fluctuations
in the major Indian stock exchanges challenges these claims, 
while at the same time, providing strong support to the universality
for the ``inverse cubic law'' which had previously only been 
seen in developed markets.
\vspace{-0.5cm}
\section{Distribution of Trading Volume and Number of Trades}
\label{sec:5}
\begin{figure}[tbp]
\centering
\includegraphics[width=0.49\linewidth,clip]{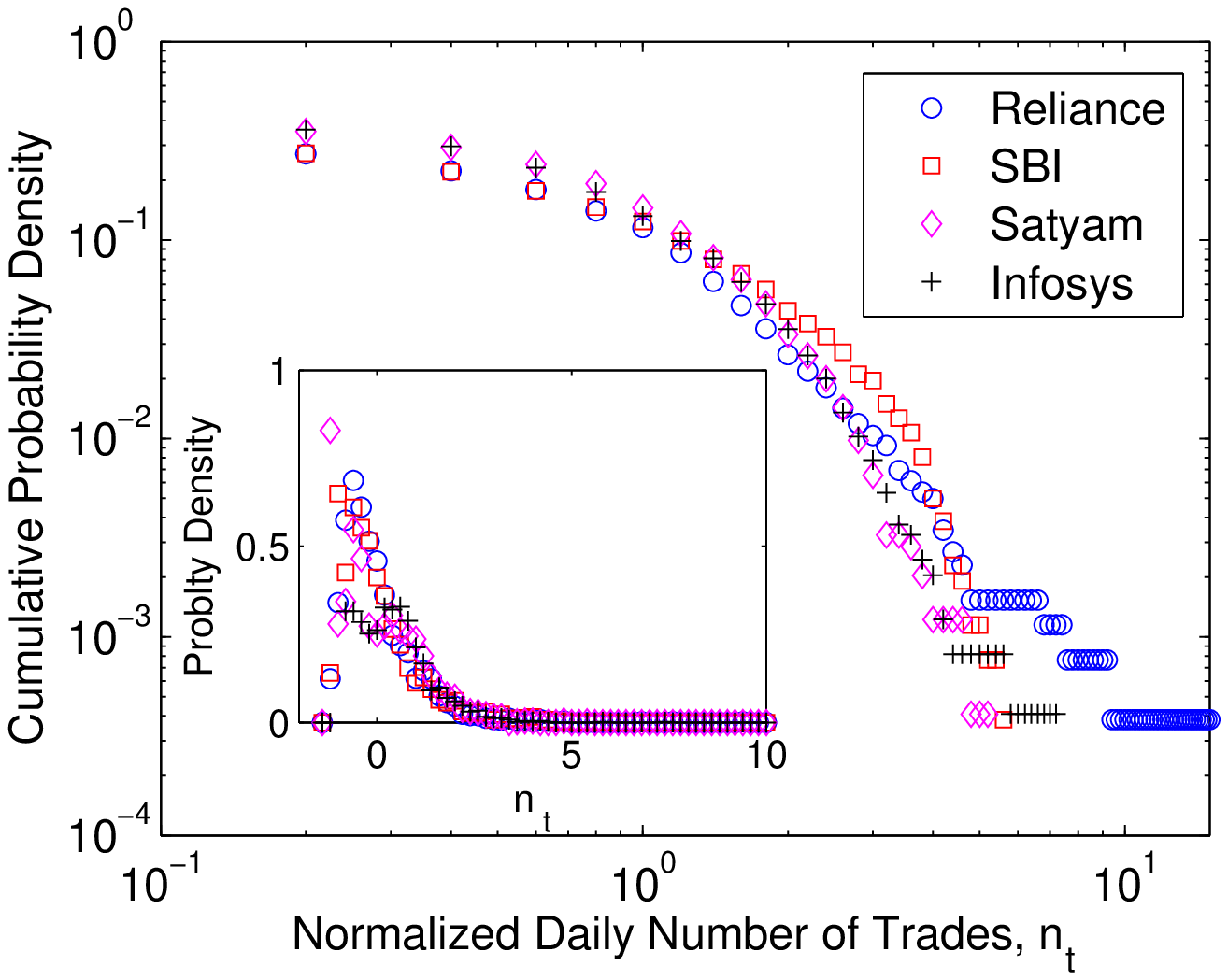}
\includegraphics[width=0.49\linewidth,clip]{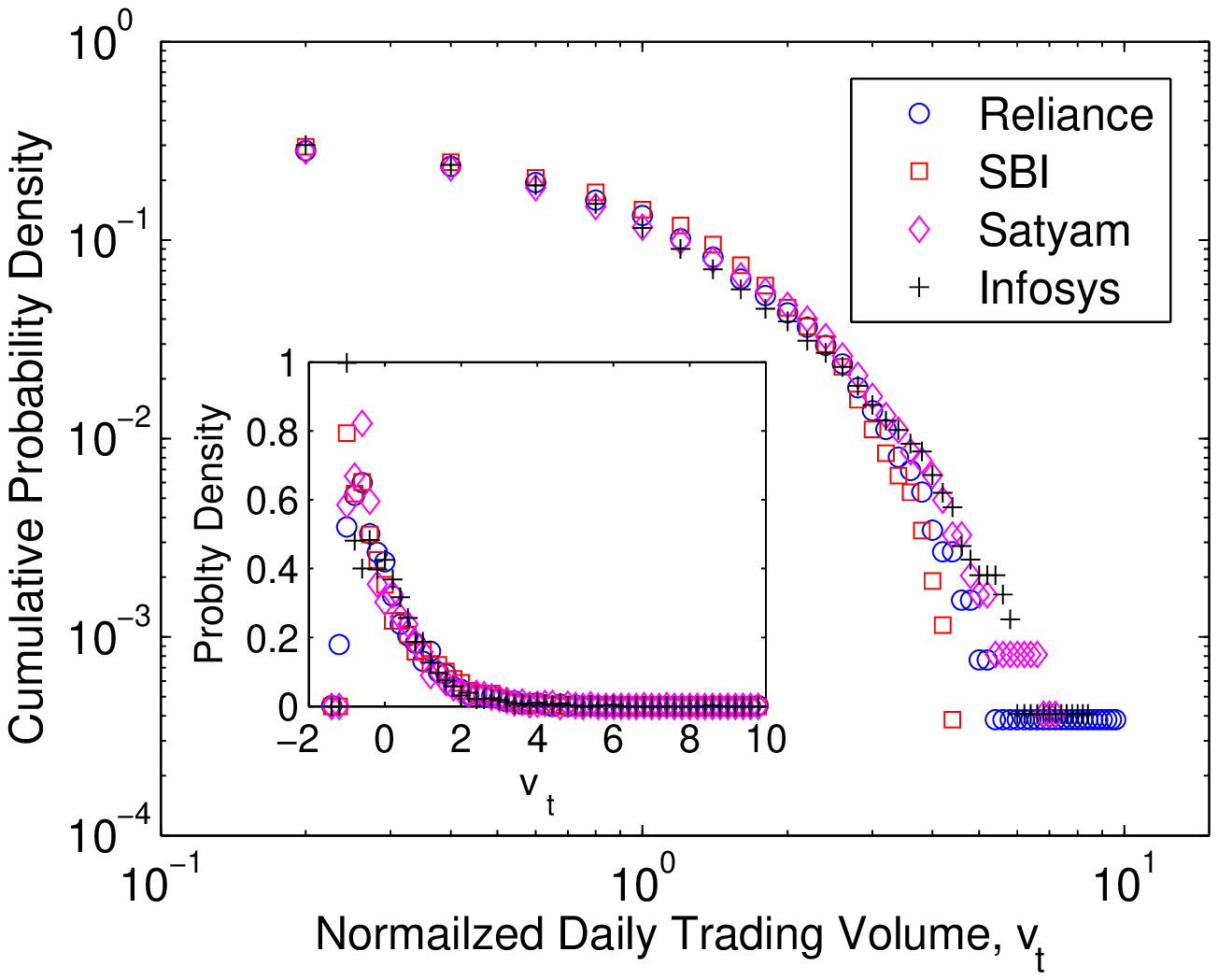}\\
\caption{Cumulative distribution of the number of trades (top left) 
and the volume of shares traded (top right) in a day for four stocks
at BSE between July 12, 1995 and Jan 31, 2006.} 
\label{dailytradevol}       
\vspace{-0.15cm}
\end{figure}

\begin{figure}[tbp]
\centering
\includegraphics[width=0.48\linewidth,clip]{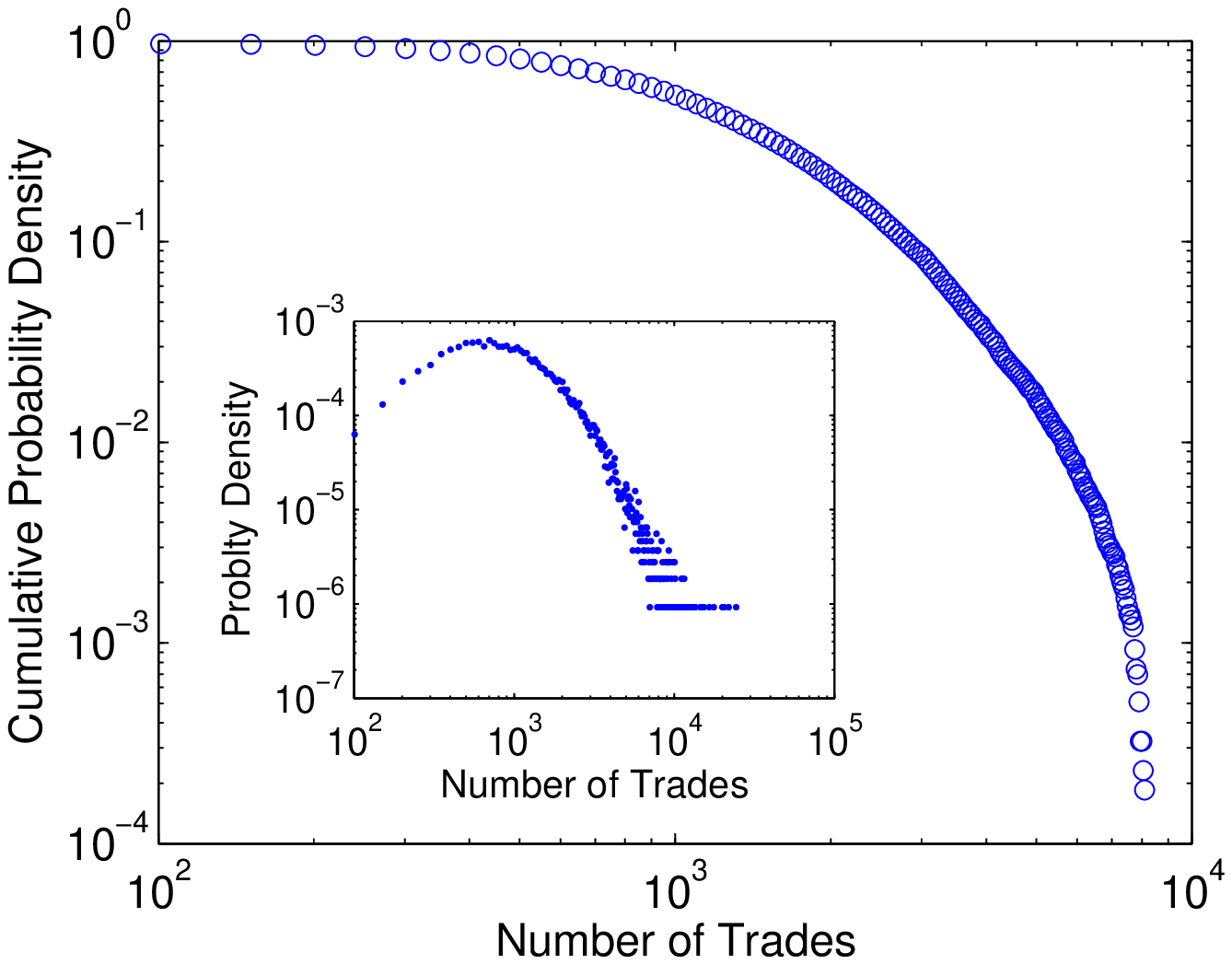}
\includegraphics[width=0.48\linewidth,clip]{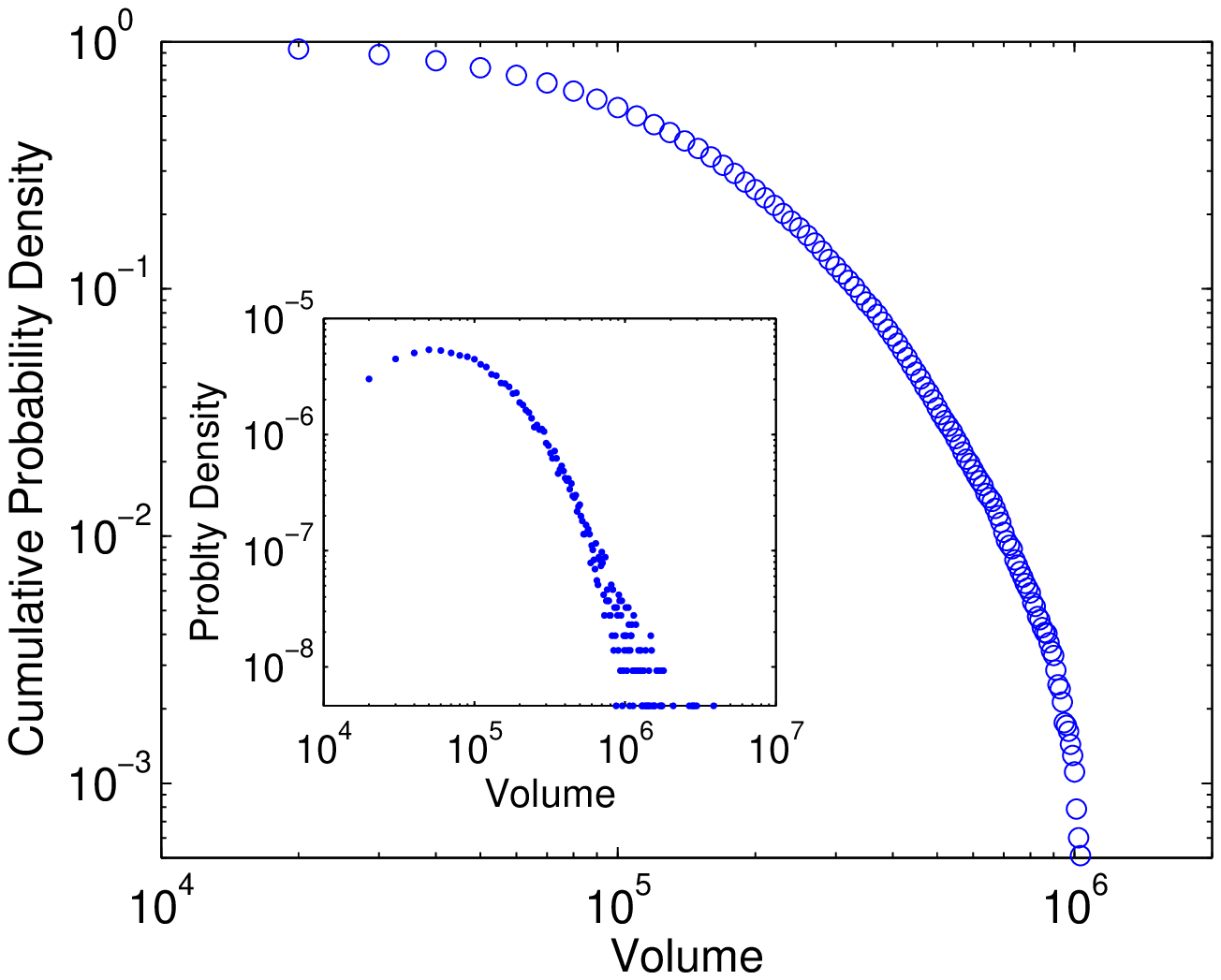}
\includegraphics[width=0.48\linewidth,clip]{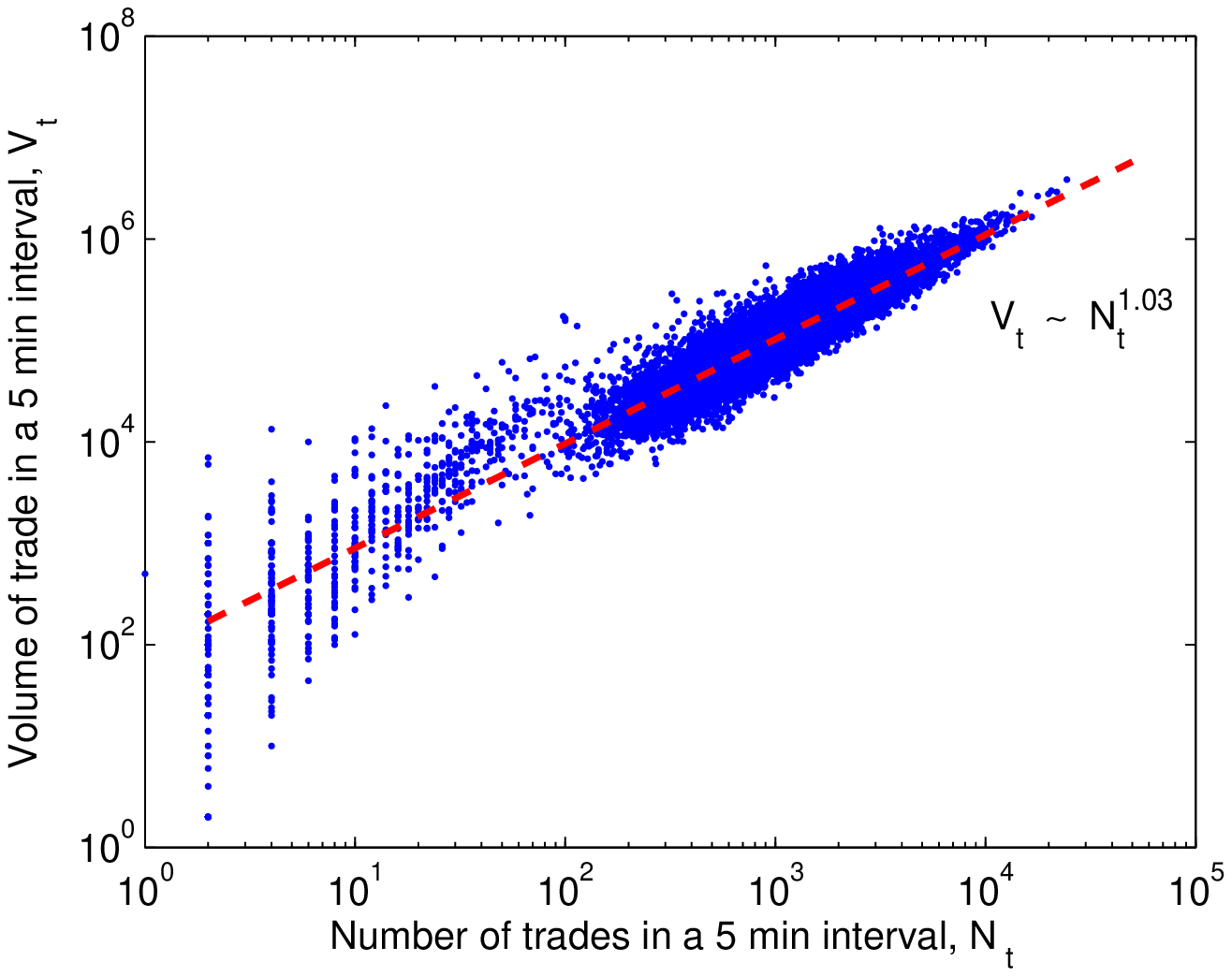}\\
\caption{Cumulative distribution of the number of trades (top left) 
and the volume of shares traded (top right) for a particular stock 
(Reliance) in 
5-minute intervals at NSE between Jan 1, 2003 to March 31, 2004. The bottom
figure shows an almost linear relation between the number of trades
in a 5-minute interval and the corresponding trading volume. The broken
line indicates the best fit on a doubly logarithmic scale.}
\label{ticktradevol}       
\vspace{-0.15cm}
\end{figure}

\vspace{-0.25cm}
Besides the price of stocks, one can also measure market activity by 
looking at the trading volume (the number of shares traded),
$V (t)$, and the number of trades, $N (t)$. To obtain the corresponding
distributions, we normalize these variables by
subtracting the mean and dividing by their standard deviation, such that,
$v =\frac{V -\langle V  \rangle}{\sqrt{\langle V^{2} \rangle - \langle V \rangle^2}}$ and
$n =\frac{N -\langle N \rangle}{\sqrt{\langle N^{2} \rangle - \langle N \rangle^2}}$ . 
Fig.~\ref{dailytradevol} shows the distribution of these two quantities
for several stocks, based on daily data for BSE. As is evident, the
distribution is very similar for the different stocks, and the nature
of the decay is significantly different from a power law.
To better characterize the distribution, we have also looked at the
intra-day distributions for volume and number of trades, based on
high-frequency data from NSE. Fig.~\ref{ticktradevol} shows the
distributions of the two quantities for trading conducted on a particular 
stock in 5 minute intervals. Analysis of data for other stocks 
show qualitatively similar results. As is clear, both of these
distributions are non-monotonic, and are suggestive of a log-normal form.
The fact that these distributions are very similar to each other is not
surprising in view of the almost linear relationship between the two
(Fig.~\ref{ticktradevol}, bottom).
This supports previous observation in major US stock markets that
statistical properties of the number of shares traded and the number
of trades in a given time interval are closely related \cite{gopikrishnan00}.

For US markets, power law tails have been reported for the distribution
of both the number of trades \cite{plerou00} and the volume \cite{gopikrishnan00}.
It has also been claimed that these features are observed on the Paris
Bourse, and therefore, these features are as universal as the
``inverse cubic law'' for price returns distribution \cite{gabaix03}. 
However, analysis
of other markets, e.g., the London Stock Exchange \cite{farmer04} have failed
to see any evidence of power law behavior. Our results confirm the latter
assertion that the power law behavior in this case may not be universal, 
and the particular form of the distribution of these quantities may be 
market specific.

\vspace{-0.5cm}
\section{Correlated Stock Movement in the Indian Market}
\label{sec:6}
\vspace{-0.25cm}
As indicated in a previous section, we now return to look at the
cross-correlation among price movements. The data that we analyze 
for this purpose consists of 2255 daily returns each for 45 stocks.
We divide this data into $M$ overlapping windows of width $T$, i.e., each 
window contains $T$ daily returns. The displacement between two
consecutive windows is given by the 
window step length parameter $\delta t$. In our study,
$T$ is taken as six months (125 trading days), while $\delta t$ is taken
to be one
month (21 trading days). 
The correlation between returns for stocks $i$ and $j$ is calculated as
\begin{equation}
C_{ij}=\langle r_{i} r_{j} \rangle - \langle r_{i}\rangle \langle r_{j} \rangle,
\label{correlation}
\end{equation}
where $\langle \ldots \rangle$ represents the time average within a window.
The resulting correlation matrices, $C$, can be analysed to get further
understanding of the relations between movements of the different stocks.

\begin{figure}[tbp] \centering
\includegraphics[width=0.49\linewidth,clip]{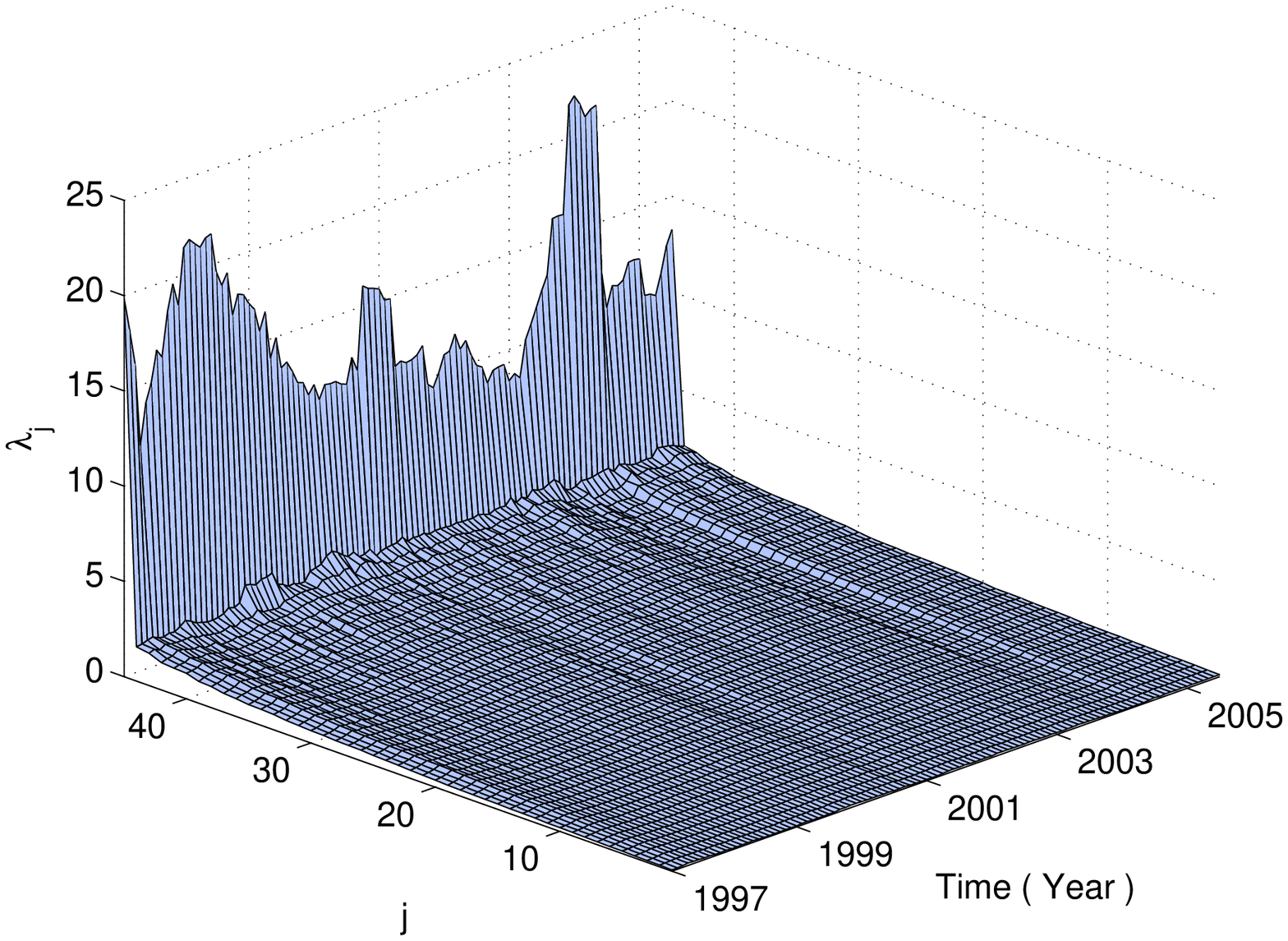}
\includegraphics[width=0.49\linewidth,clip]{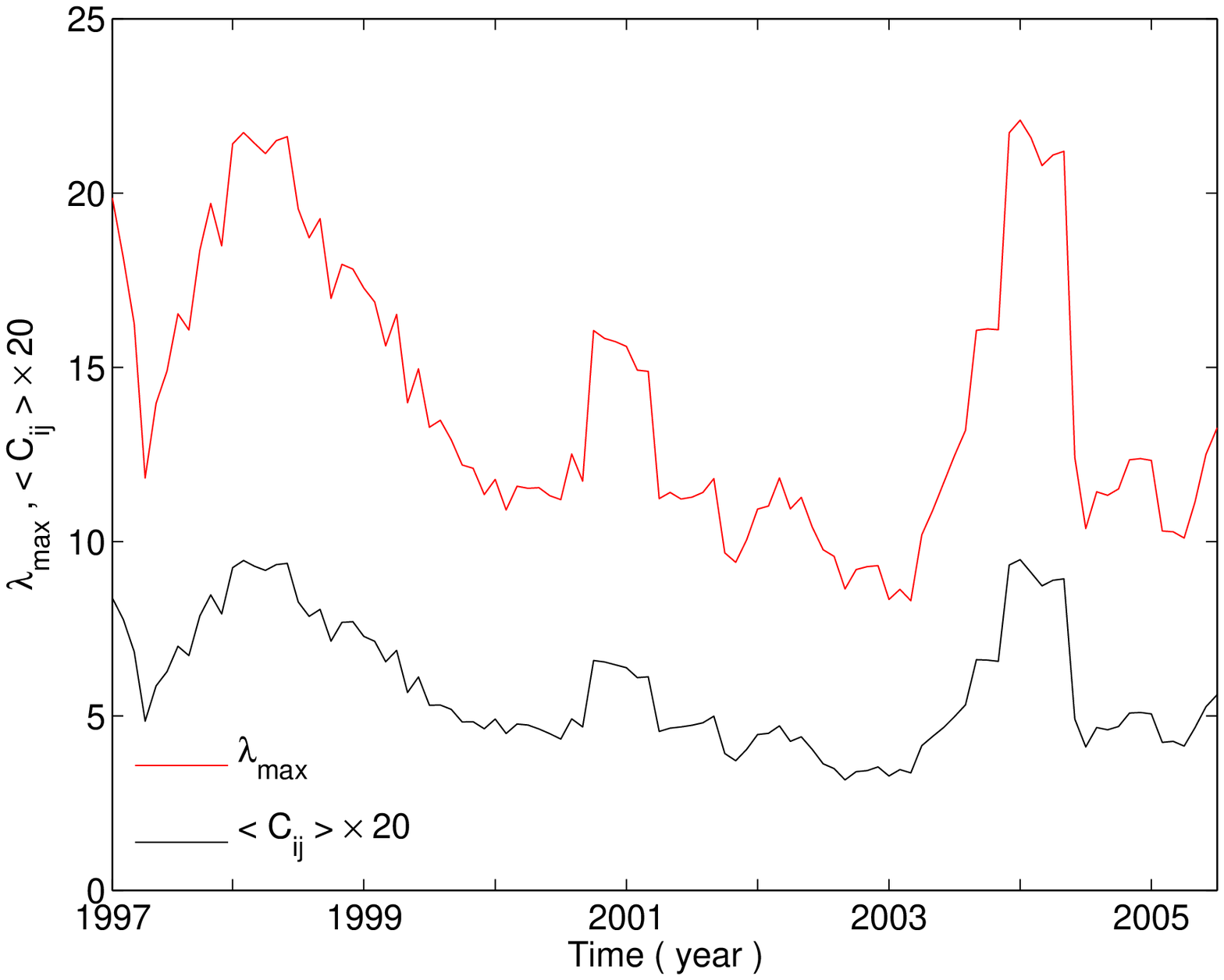}
\caption{(Left) The eigenvalues, sorted in descending order,
for the correlation matrices of daily price returns for 45 stocks in NSE, 
across time. (Right) The variation across time of the largest eigenvalue 
$\lambda_{max}$ of the correlation matrices
and the average correlation $\langle C \rangle$.
The window size chosen for calculating correlation is 6 months 
and the window is shifted
in steps of 21 days. The total period is from Jan 1997 to Dec 2005.}
\label{eigenval}       
\vspace{-0.15cm}
\end{figure}

We now look at the eigenvalues of $C$ which contain significant information
about the cross-correlation structure \cite{plerou02}.
Fig.~\ref{eigenval} (left) shows the eigenvalues of $C$ as a function
of time. It is clear that the majority of these are very close to zero
at all times. The largest eigenvalues contain almost all information 
about the market, which is evident from Fig.~\ref{eigenval} (right).
This shows the variation of the average correlation coefficient,
as well as the largest eigenvalue $\lambda_{max}$, with time. The two
are strongly correlated, indicating that $\lambda_{max}$ captures
the behavior of the entire market. Our results indicate that the
Indian market is highly correlated, as indicated by the strong
cross-correlations among the most traded stocks.

\vspace{-0.5cm}
\section{Conclusions}
\vspace{-0.25cm}
In this paper, we have examined the statistical properties of trading in
the two largest Indian financial markets, 
BSE and NSE. Using both low-frequency (daily) and high-frequency 
(tick-by-tick), we demonstrate that the price return cumulative distribution 
has long tails, consistent with a power
law having exponent close to 3. This lends strong support to the claim
that the price return distribution has an universal form across
different markets, namely, the
``inverse cubic law''.
On the other hand, the distributions for volume and number of trades appear to
be log-normal, the two quantities being almost linearly related. 
We also find the market index fluctuation distribution to have the same
form as the distribution of individual stock price returns. This implies
that stocks in the Indian market are highly correlated. We verify that
this is indeed the case with
a cross-correlation analysis of most of the frequently traded stocks 
in the Indian market.

\vspace{0.1cm}
{\small {\bf Acknowledgements:} We are grateful to M.~Krishna for 
invaluable assistance in obtaining and analyzing
the high-frequency NSE data. We thank S.~Sridhar
and N.~Vishwanathan for technical assistance in arranging the data,
and J.-P. Onnela for helpful discussions.}

\vspace{-0.4cm}

\printindex
\end{document}